\documentclass[12pt,a4paper]{article}

\begin{document}

\title{Neurino propagation in matter using the wave packet approach}
\author{J.T. Peltoniemi$^{1,2}$\thanks{juha.peltoniemi@oulu.fi} \  and 
 V. Sipil\"{a}inen$^1$\thanks{ville.sipilainen@helsinki.fi} \\
  $^1$\emph{Department of Physics, FIN-00014 University of Helsinki, Finland}\\
  $^2$\emph{Centre for Underground Physics in Pyh\"asalmi}\\
\emph{Sodankyl\"a Geophysical Observatory}\\
\emph{FIN-90014 University of Oulu, Finland}
    }
\date{April 16, 2000}
\maketitle

\begin{abstract}
We study the oscillations and conversions of relativistic neutrinos 
propagating in matter of variable density using the wave packet 
formalism.
We show how the oscillation and coherence lengths are modified 
in comparison with the case of oscillations in
vacuum. 
Secondly, we demonstrate how the equation of motion for two
neutrino flavors can be formally solved for
almost arbitrary density profile.
We calculate finally how the use of wave packets alters the
nonadiabatic level crossing probabilities. 
For the most common physical environments the corrections 
due to the width of the wave packet do not lead to
observable effects.

\end{abstract}
\hspace{1cm}PACS: 03.65.-w, 14.60.Pq

\vspace{-15cm}
\hspace{12.5cm}UO-Report 130/2000

\newpage
\section{Introduction}
Recent experimental discoveries 
\cite{5} seem to suggest that
neutrino oscillations \cite{1,2,3,4} really exist. 
Since the observation of neutrino oscillation may provide valuable
information on the basic properties of neutrinos, 
e.g.\ masses and mixing angles,
it is important to know the underlying 
physics also on the conceptual level.

It was pointed out e.g.\ in Refs.\ 
\cite{6} and \cite{7} that the standard
quantum mechanical treatment of 
neutrino oscillations using plane waves 
\cite{4,8} is not completely satisfactory for
many reasons. 
The wave packet approach \cite{6},\cite{9}--\cite{15}
provides a more physical picture which is particularly adapted 
for describing phenomena localized in space and time.
This formalism also elegantly accounts for the loss of
coherence by the separation of wave packets.
However, 
some authors (e.g.\ \cite{7,17,19}) have been skeptical about the use of 
wave packets, and Refs.\ \cite{19,20} conclude that the concept of wave
packet is unnecessary for all the relevant physical cases. 
A bunch of other 
methods has also been discussed \cite{7,17,19,16,18} (to name a few). 

In this paper we will consider neutrino 
oscillations and other phenomena in matter.
We have decided to use wave packets because 
the calculations of the kind presented here 
have never been carried out before. 

We will first focus on the equation of motion 
for neutrinos propagating in matter of variable 
density. Generally the equation is modified in
matter due to coherent forward scatterings
\cite{21,22,23,24}, and an exact solution can
be found only for few special cases.
Here we address one special case where the
density of matter changes slowly enough, so that
the situation is said to be adiabatic.
Now the eigensolutions for the equation of motion 
are found trivially, for arbitrary number of relativistic neutrinos. 
To describe neutrino oscillations in matter, we
apply the method of Ref.\ \cite{14} in connection 
with these solutions. 
We can express the results formally using effective
oscillation and coherence lengths which are
not local quantities anymore.

Generally neutrinos may propagate nonadiabatically, and solving the
complete equation of motion even for two neutrino flavors is usually far from
trivial (see e.g.\ \cite{25,26,27} for two specific density profiles). In this
paper we show that the solution for ``arbitrary'' density profile can be
constructed by using infinite integral series. Some supplementary calculations
which may be of formal interest are enclosed in Appendix B\@.
Nonadiabaticity is also
related to the fact that so-called level crossings (or hoppings) between matter
``eigenstates'' take place (\cite{24,28} and references therein). Our calculations
show that the level crossing probabilities are modified when neutrinos are
described by wave packets.

\section{Neutrino propagation in matter in the adiabatic limit}

\subsection{Equation of motion}

Barring the details of the production and detection mechanisms
of neutrinos, we we will focus on the propagation.
Traditionally the propagation of neutrinos is modeled 
by solving the relativistic Schr\"{o}dinger equation with the 
effective Hamiltonian 
\begin{equation}
    \hat{H} = \sqrt{\hat{p}^{2}+m^2 } + V(\hat{x},t) 
     \approx \hat{p} + \frac{1}{2}m^2 \hat{p}^{-1} + V(\hat{x},t)\ ,    
\label{e1} \end{equation}
where $\hat{p}$ is the momentum operator, $m$ is the neutrino mass matrix and
$V(x,t)$ is a semiclassical potential due to the presence of medium. 
In this work we assume that $V(x)$ does not depend on time,
and that it is a matrix diagonal in the weak interaction basis or
flavor basis. This is well justified when considering neutrinos
from the Sun, from supernovae, or on Earth. On the other
hand, this assumption excludes neutrinos in the early
universe that must be treated otherwise.

Instead of decomposing the wave function in momentum space,
we prefer to consider the Fourier transform in energy space,
\begin{equation}
    \psi(x,t) = 
\frac{1}{\sqrt{2\pi}} \int dE e^{-iEt} 
\psi(x,E)\ . \label{2}
\end{equation}
This leads to a more physical picture when the Hamilton operator
is a function of $x$. Particularly, now the energy $E$ is a constant,
while the momentum $p$ is a function of $x$.
Throughout this work we will assume the propagation to
be a one-dimensional phenomenon.

The resulting time independent Schr\"{o}dinger equation
can be written in the relativistic limit as
\begin{equation}\label{tis}
i\partial_{x}\psi(x,E) = 
\left(-E+\frac{m^{2}}{2E} + V(x) \right)
\psi(x,E) + O\left(\frac{m^4+m^{2}EV}{E^3} \right)\psi(x,E)\ , 
\end{equation}
where we have also assumed that $E \gg V$.
The right-hand side can be locally diagonalized.
We call the respective eigenvectors as local matter
eigenstates, and the respective eigenvalues
can be written as
\begin{equation}
p_a (x,E)
\equiv
E - \frac{\mu_a^{2}(x,E)}{2E},
\end{equation}
where we introduced the effective mass $\mu_a(x,E)$
that simplifies the notation.
In this paper
the Greek indices refer to the flavor basis and the Latin indices to the matter
basis. 
In the adiabatic limit, or for negligible mixing, 
Eq.\ (\ref{tis}) can be written as
\begin{equation}
i\partial_{x}\psi_{a}(x,E)\approx 
\left(-E+\frac{\mu_{a}^{2}(x,E)}{2E}\right)
\psi_{a}(x,E)
= -p_{a}(x,E)\psi_{a}(x,E)\ ,   \label{3}
\end{equation}
Outside the adiabatic limit the complete equation of motion includes 
also nondiagonal terms (compare to Eq.\ (11) of Ref.\
\cite{29}). 

  From now on we follow the treatment presented in Ref.\ \cite{14}.
Eq.\ (\ref{3}) can be solved easily, and one has
\begin{equation}
     \psi_{a}(x,t)=\frac{1}{\sqrt{2\pi}}\int
dE\exp\left[i\int_{0}^{x}dx'p_{a}(x',E)-iEt\right]\psi_{a}(0,E)\ ,
\end{equation}
where Eq.\ (\ref{2}) was used. The initial value 
of $x$ is put to zero since we consider neutrinos
produced at the origin. 

\subsection{Wave packet solutions}

As  initial condition, we assume that the neutrino
wave function is a linear combination of
matter eigenstates,
\begin{equation}
\psi(0,E) = \sum_a C_a(E) \psi_a(0,E).
\end{equation}
In the relativistic limit the coefficients $C_a$ match
the respective row in the mixing matrix,
$C_a(x,E) = U_{\alpha a}^{\ast}(x,E)$, but for less relativistic
neutrinos correction factors related to energy
conservation must be taken into account \cite{30}.
We assume further that each component of the wave function
$\psi_{a}(0,E)$ has a Gaussian form
\begin{equation}
\psi_{a}(0,E)=(2\pi\sigma_{EP}^{2})^{-1/4} 
\exp\left[-\frac{(E-E_{a})^{2}}{4\sigma_{EP}^{2}}\right]\ ,    \label{6}
\end{equation}
where $E_{a}$ is the average energy of the corresponding matter eigenstate and
$\sigma_{EP}$ is the energy width related to the production process fulfilling the
uncertainty relation $\sigma_{EP}\sigma_{tP}=1/2$.

In the adiabatic limit we obtain for the partial wave packet the 
solution 
\begin{equation}
    \psi_{a}(x,t) = C_1
\int dE \exp \left[ i\int_0^x dx'p_a (x',E)-iEt-
\frac{(E-E_a)^2}{4\sigma_{EP}^2 }\right] ,
\end{equation}
where $C_1$ is a numerical factor.
We assume that the wave packet is sufficiently narrow in energy
space, $\sigma_{EP} \ll E_a$. This helps us to integrate the above 
integral and we also avoid considering the negative energies. 
This assumption is justified e.g. for solar neutrinos \cite{12,24}.
One can now expand the momentum as
\begin{equation}
p_{a}(x,E)\approx
p_{a}(x,E_{a})+
\frac{\partial p_{a}}{\partial E_{a}} (E-E_{a})
=p_{a}(x,E_{a})+\frac{E-E_{a}}{v_{a}(x,E_a)}\ ,
\end{equation}
where $v_{a}(x,E)$ is the
group velocity of each wave packet. 
Note that within this framework the group velocity
depends on $x$ via the potential $V(x)$,
unlike in some alternative works.
Writing $\sigma_{EP}\approx v_{a}(0)\sigma_{pP}=v_{a}(0)/(2\sigma_{xP})$,
one has after a simple integration 
\begin{equation}
\psi_{a}(x,t)=C_2 \exp\left[i\int_{0}^{x}dx'p_{a}(x',E_{a})-iE_{a}t
-\frac{v_{a}^{2}(0)}{4\sigma_{xP}^{2}}
\left(\int_{0}^{x}\frac{dx'}{v_{a}(x')}-t
\right)^{2}\right]\ ,  \label{lisays}
\end{equation}
where $v_{a}(0)$ is the group velocity at the origin and
$\sigma_{pP}(\sigma_{xP})$ is the momentum (spatial) width related to the
production process. 
Eq.\ (\ref{lisays}) implies that the wave 
function of a flavor neutrino $\nu_{\alpha}$ is given 
by \cite{14}
\begin{equation}
    |\nu_{\alpha}(x,t)\rangle =C_2 \sum_{a}
U_{\alpha a}^{\ast}(0)
\exp\left[i\int_ {0}^{x}dx'p_{a}(x',E_{a})-iE_{a}t
-\frac{v_{a}^{2}(0)}{4\sigma_{xP}^{2}}
\left(\int_{0}^{x}\frac{dx'}{v_{a}(x')}
-t\right)^{2}
\right]|\nu_{a}\rangle\ ,
\label{10}
\end{equation}
where $U(x)$ represents the effective mixing matrix and the 
states $|\nu_{a}\rangle$ are orthonormal.

To get out relevant physics we need to know the size of the wave packet.
This is very non-trivial and lots of rather confusing estimates have 
been presented. Here we assume that the width of the neutrino wave is 
mostly related to the spatial details of the production process.
In principle it also depends on temporal properties, like the stability
of the state producing the neutrino, but this is relevant only
for very short-lived particles.
It has been estimated (\cite{12,24} and relevant references therein)
that $\sigma_{xP}$ is of the order 
$10^{-9}\ \mbox{m}$ for solar neutrinos,
$\sigma_{xP}\sim 10^{-11}\ \mbox{m}$ at the neutrino sphere
of a supernova, and
$\sigma_{xP}\sim 10^{-6}\ \mbox{m}$ for reactor neutrinos, for example.

\subsection{Observation of neutrino oscillation}

The quantum state of the neutrino is  measured by
an appropriate reaction. Assuming the respective process
being $\nu_\alpha + X \to $ (something visible), we can relate the 
quantum mechanical uncertainty of the detection process to
the quantum mechanical state of the particle $X$
before the collision.
Here we assume that the relevant wave functions are 
Gaussians, centered 
at a distance $L$ from the origin 
(source), with a spatial width $\sigma_{xD}$.
Hence the detection can be described by  
\begin{equation}
    |\nu_{\beta}(x-L)\rangle =C_{2}'\sum_{a}
U_{\beta a}^{\ast}(L)
\exp\left[i\int_ {L}^{x}dx'p_{a}(x',E_{a})
-\frac{v_{a}^{2}(L)}{4\sigma_{xD}^{2}}
\left(\int_{L}^{x}\frac{dx'}{v_{a}(x')}\right)^{2}
\right]|\nu_{a}\rangle\ ,
\label{11}
\end{equation}
where $v_{a}(L)$ is the group velocity at $L$.
Note that this is independent of time \cite{14}.

We emphasize that the spatial uncertainty, $\sigma_{xD}$,
arises for purely quantum-mechanical reasons.
One may be apt to believe that it is at most of the order
of atomic distances, i.e. $\sigma_{xD}\sim 10^{-9}\ \mbox{m}
-10^{-10}\ \mbox{m}$ or less.
In reality
there are also other instrumental
uncertainties related to
the resolution or dimension of the detector.
These can be accounted for by taking an
average over the relevant length scale.

The amplitude of the
process $\nu_{\alpha}\rightarrow\nu_{\beta}$ is now simply
\begin{eqnarray}
     A_{\alpha\beta}(L,T) &=&
\int dx\langle\nu_{\beta}(x-L)|\nu_{\alpha}(x,T)\rangle
\nonumber  \\
& = &
C_{2}'' \sum_{a}U_{\alpha a}^{\ast}(0)U_{\beta a}(L)
\int
dx \exp \left[i\int_{0}^{L}dx'p_{a}(x',E_{a})-iE_{a}T \right.
\nonumber  \\ && \left.
\mbox{} - \frac{v_{a}^{2}(0)}{4\sigma_{xP}^{2}}
\left(\int_{0}^{x}\frac{dx'}{v_{a}(x')}-
T\right)^{2} - \frac{v_{a}^{2}(L)}{4\sigma_{xD}^{2}}
\left(\int_{L}^{x}\frac{dx'}{v_{a}(x')}\right)^{2}
\right]\ .
   \label{12}
\end{eqnarray}
The first two terms in the exponential
are independent of $x$, but the integration of the two other terms is far from
obvious. Notice that the analytical form of $v_{a}(x)$ is unknown in general, and
that we have not even chosen any specific density profile so far. It turns out,
however, that the integral can be evaluated
with a saddle point method to a sufficiently good approximation.

After a straightforward, but rather lengthy calculation, presented 
in Appendix A, we obtain the amplitude 
 (with Eqs.\ (\ref{17}) and (\ref{uusi}))
\begin{equation}\label{amp}
      A_{\alpha\beta}(L,T) =C_3 \sum_{a}U_{\alpha a}^{\ast}(0)U_{\beta
a}(L)\exp\left[i\int_{0}^{L}dxp_{a}(x,E_{a})-iE_{a}T
-\frac{1}{4}\frac{\left(T-\int_{0}^{L}\frac{dx}{v_{a}(x)}\right)^{2}}
{\left(\frac{\sigma_{xP}}{v_{a}(0)}\right)^{2}+
\left(\frac{\sigma_{xD}}{v_{a}(L)}\right)^{2}}\right]\ .
\end{equation}
The respective probability of the process 
$\nu_{\alpha}\rightarrow\nu_{\beta}$ is
then
\begin{eqnarray}    \label{23}
P_{\alpha\beta}(L,T)&=&|A_{\alpha\beta}(L,T)|^{2}
\nonumber \\
&\sim&
\sum_{a,b}U_{\alpha a}^{\ast}(0)
U_{\beta a}(L)U_{\alpha b}(0)U_{\beta b}^{\ast}(L) e^{G(L,T)}\ ,
\end{eqnarray}
where
\begin{eqnarray}   
G(L,T) &=&
i\int_{0}^{L}dx \left(p_{a}(x,E_{a})-p_{b}(x,E_{b})\right)
-i(E_{a}-E_{b})T
 \nonumber \\ && \mbox{} 
-\frac{1}{4\chi_{a}}
\left(T-\int_{0}^{L}\frac{dx}{v_{a}(x)}\right)^{2}
-\frac{1}{4\chi_{b}}
\left(T-\int_{0}^{L}\frac{dx}{v_{b}(x)}\right)^{2}
\end{eqnarray}
with \begin{equation}
\chi_{a}\equiv\left(\frac{\sigma_{xP}}{v_{a}(0)}\right)^{2}
+\left(\frac{\sigma_{xD}}{v_{a}(L)}\right)^{2}\ .
\end{equation}

We still have to perform an integration over time in $P_{\alpha\beta}(L,T)$ since
we are mainly interested in $P_{\alpha\beta}(L)$ \cite{6,14}. A straightforward
Gaussian integration of the exponential of Eq.\ (\ref{23}) yields
\begin{eqnarray}  
\int dT e^{G(L,T)} & \sim &
\exp \left\{ i\int_{0}^{L}dx(p_{a}(x,E_{a})-p_{b}(x,E_{b}))
\right\} 
\nonumber \\ &&
\exp \left\{
-\frac{1}{4\chi_{a}}\left(\int_{0}^{L}\frac{dx}{v_{a}(x)}\right)^{2}
-\frac{1}{4\chi_{b}}\left(\int_{0}^{L}\frac{dx}{v_{b}(x)}\right)^{2}\right\}
\nonumber \\ &&
  \!
\times\exp\left\{\frac{\chi_{a}\chi_{b}}
{\chi_{a}+\chi_{b}}\left[-(E_{a}-E_{b})^{2}+\frac{1}{4}
\left(\frac{1}{\chi_{a}}
\int_{0}^{L}\frac{dx}{v_{a}(x)}+
\frac{1}{\chi_{b}}\int_{0}^{L}\frac{dx}{v_{b}(x)}\right)^{2}
\right. \right. \nonumber \\ && \left. \left.
-i(E_{a}-E_{b})\left(\frac{1}{\chi_{a}}\int_{0}^{L}\frac{dx}{v_{a}(x)}
+\frac{1}{\chi_{b}}\int_{0}^{L}\frac{dx}{v_{b}(x)}\right)\right]\right\}
 \label{25}
\end{eqnarray}
 without the prefactor, which is a constant in the relativistic limit. A detailed
analysis renders Eq.\ (\ref{25}) to 
\begin{eqnarray}
\int dT e^{G(L,T)} & \sim &
\exp\left\{i\int_{0}^{L}dx
(p_{a}(x,E_{a})-p_{b}(x,E_{b}))-i(E_{a}-E_{b})L
\right.  \nonumber \\ && \left. 
-\frac{1}{8\sigma_{x}^{2}}
\left[\int_{0}^{L}dx(v_{a}(x)-v_{b}(x))\right]^{2}
-\frac{(E_{a}-E_{b})^{2}}{8\sigma_{p}^{2}}\right\}\
,  \label{26}
\end{eqnarray}
where $\sigma_{x}^{2}\equiv\sigma_{xP}^{2}+\sigma_{xD}^{2},\ \sigma_{p}\equiv
1/(2\sigma_{x})$ and the relativistic limit is once again considered.

In order to simplify Eq.\ (\ref{26}), we approximate (compare to Ref.\ \cite{14})
\begin{equation}
   E_{a}\approx E_0+\xi\frac{\mu_{a}^{2}(0)}{2E_0}\ ,  \label{27}
\end{equation}
where $\mu_{a}(0)$ is the effective mass at the origin,
$E_0$ is the central energy in the limit of zero neutrino masses and 
$\xi$ is a parameter of
order unity, related to the energy-momentum conservation of the 
production process.
The corresponding momentum is given by
\begin{equation}
   p_{a}(x,E_{a})
\approx 
E_{a}-\frac{\mu_{a}^{2}(x)}{2E_{a}}
\approx
E_0+\xi\frac{\mu_{a}^{2}(0)}{2E_0}-\frac{\mu_{a}^{2}(x)}{2E_0}\ .  
\label{28} \end{equation}
Notice that $E_{a}$ is a constant, but $p_{a}(x,E_{a})$ depends on $x$, thus this
approximation does not contradict the chosen convention. One may also point out
that the initial dependence $\mu_{a}=\mu_{a}(x,E_{a})$ has reduced to
$\mu_{a}=\mu_{a}(x,E_{0})$. Using Eqs.\ (\ref{27}) and (\ref{28}), Eq.\ (\ref{26})
gives
\begin{equation} 
\int dT e^{G(L,T)} \sim
\exp\left[-\frac{i}{2E_0}\int_{0}^{L}dx\Delta\mu^2_{ab}(x)
-\frac{1}{8\sigma_{x}^{2}}\left(\int_{0}^{L}dx\Delta
v_{ab}(x)\right)^{2}-\frac{(E_{a}-E_{b})^{2}}{8\sigma_{p}^{2}}\right]\ ,
\end{equation}
where $\Delta\mu^2_{ab}(x)\equiv\mu_{a}^{2}(x)-\mu_{b}^{2}(x),\ \Delta v_{ab}(x)\equiv
v_{a}(x)-v_{b}(x)$ and the last term has not been altered on purpose. 

Combining the above results we obtain a compact formula for
the probability to observe a neutrino 
in a state $\nu_\beta$ at a distance $L$
(with the correct normalization $\sum_{\beta}P_{\alpha\beta}(L)=1$),
\begin{eqnarray}
   P_{\alpha\beta}(L) & = & \sum_{a,b}U_{\alpha a}^{\ast}(0)U_{\beta
a}(L)U_{\alpha b}(0)U_{\beta b}^{\ast}(L) \nonumber  \\
   &  &  \times\exp\left[-2\pi
i\frac{L}{L_{ab}^{osc}(L)}-\left(\frac{L}{L_{ab}^{coh}(L)}\right)^{2}
-\frac{(E_{a}-E_{b})^{2}}{8\sigma_{p}^{2}}\right]\
,  \label{30}
\end{eqnarray}
where the effective oscillation and coherence lengths are defined by 
\begin{equation}
  L_{ab}^{osc}(L)\equiv\frac{4\pi E_0 L}{\int_{0}^{L}dx\Delta\mu^{2}_{ab}(x)}
\ ,\ \ \ \
L_{ab}^{coh}(L)\equiv\frac{2\sqrt{2}\sigma_{x}L}{\left|\int_{0}^{L}dx\Delta
v_{ab}(x)\right|}\ .   \label{31}
\end{equation}
The problem has reduced to computing the integrals
over the effective mass and group velocity differences along
the neutrino path.

Let us comment on the following issues in connection with Eqs.\ (\ref{30}) and
(\ref{31}):
\begin{enumerate}
\item The fact that the effective mixing matrices depend on the production and
detection location is one part of the well-known MSW effect \cite{21,31}.
\item Compared to the usual oscillation and coherence lengths, 
$L^{osc}=4\pi E_0 /\Delta_{0},\ L^{coh}=
2\sqrt{2}\sigma_{x}/|\Delta v_{0}|$ 
(where
$\Delta_{0}(\Delta v_{0})$ is the mass squared (velocity) difference in vacuum),
we see that the corresponding effective lengths take into account the changes of
$\Delta\mu^{2}_{ab}(x)$ and $\Delta v_{ab}(x)$ over the whole path of propagation. In
this sense the oscillation and coherence lengths are not local quantities
anymore.
\item The physical coherence length, corresponding to a length scale where
the oscillation ceases, can be obtained by solving the equation
$L=L_{ab}^{coh}(L)$. The physical oscillation length for variable density
lacks a clear definition.
\item The last term of the exponential in Eq.\ (\ref{30}) is related to the energy
conservation within the uncertainty $\sigma_{p}$. Its physical meaning is easy to
understand: if e.g.\ $|E_{1}-E_{2}|\gg\sigma_{p}$, only one of the states
$\nu_{1}$ or $\nu_{2}$ is ``allowed'', i.e.\ there is no oscillation.
\item The treatment presented here is not limited to any specific density profile
if only Eq.\ (\ref{3}) is valid (i.e.\ the adiabaticity is in effect, the matter
density is not too high etc.). 
\end{enumerate} 
The discussion of Ref.\ \cite{14} is in many respects applicable also here.

We finally point out that the calculation of $\int_{0}^{L}dx\Delta v_{ab}(x)$ 
is trivial in the relativistic limit:
the group velocity is given by definition by
\begin{equation}
    v_{a}(x)=\frac{\partial E_{a}}{\partial p_{a}}\approx
1-\frac{\mu_{a}^{2}(x)}{2p_{a}^{2}(x)}
+\frac{1}{2p_{a}(x)}\frac{\partial
\mu_{a}^{2}}{\partial p_{a}}\ ,
\end{equation}
and taking into account the approximation of Eq.\ (\ref{28})
\begin{equation}
   v_{a}(x)\approx
1-\frac{\mu_{a}^{2}(x)}{2E^{2}_0}
+\frac{\partial_{E_0}\mu_{a}^{2}}{2E_0}\ .
\end{equation}
Hence 
\begin{equation}
   \int_{0}^{L}dx\Delta
v_{ab}(x)\approx\frac{1}{2E_0^{2}}(-1+E_0\partial_{E_0})
\int_{0}^{L}dx\Delta\mu^{2}_{ab}(x)\ , \label{41}
\end{equation}
i.e.\ we see that $\int_{0}^{L}dx\Delta v_{ab}(x)$ can be 
obtained easily if $\int_{0}^{L}dx\Delta\mu^{2}_{ab}(x)$ is known.

\section{Examples}
As an example we calculate the effective oscillation and coherence lengths for two
specific density profiles. For simplicity only two
neutrino flavors, $\nu_{e}$ and $\nu_{\mu}$,
are considered. 

The effective mass squared difference in matter
is given by \cite{24}
\begin{equation}
   \Delta_{m}(x)=\sqrt{(\Delta_{0}\cos
2\theta-2\sqrt{2}E_0 G_{F}N_{e}(x))^{2}+\Delta_{0}^{2}\sin^{2}2\theta}
\ , \label{32}
\end{equation}
where $\Delta_{0}$ is the mass squared difference in 
vacuum, $\theta$ is the vacuum mixing angle, $G_{F}$ is the Fermi
constant and
$N_{e}(x)$ is the number density of electrons. 
(For other neutrino flavors or exotic matter contents
$N_{e}(x)$ should be replaced by an appropriate 
combination of the particle densities of the matter.)
The resonance, where the effective mixing is maximal,
is reached at the density
\begin{equation}
    N_{e}(x_{R})=\frac{\Delta_{0}\cos 2\theta}
{2\sqrt{2}E_0 G_{F}}\ .  \label{34}
\end{equation}
The passage through the resonance is 
adiabatic when 
\begin{equation}
Q\equiv\frac{\Delta_{0}\sin^{2}2\theta}{E_0 \cos2\theta}\left|
\frac{N_{e}(x_{R})}{N_{e}'(x_{R})}\right|\gg
1\ ,   \label{33}
\end{equation}  
where $Q$ is the adiabaticity parameter.
It is also assumed that $|G_{F}N_{e}(x)|\ll E_0$.

One should notice that the matter may affect considerably the oscillation 
of neutrinos propagating in constant density. This is the case e.g.\ if
$N_e$ is very close to its resonance value and $\theta$ is small.
Then (see Eq.\ (\ref{32})) $\Delta_m \ll \Delta_0$ and the oscillation
length may be highly longer than in the vacuum case.
Since at the surface of Earth $\rho\sim 3\ \mbox{g}/ \mbox{cm}^{3}$ 
and $Y_e \approx 1/2$, and hence $G_F N_e \sim 10^{-13}$ eV, 
the effect might be observable in long baseline experiments e.g.\ for
$E_0 \sim 10\ \mbox{GeV}$, $\Delta_0 \sim 10^{-3}\ \mbox{eV}^2$ (and
small $\theta$). Also the group velocity difference, and consequently
the coherence length, can be modified for suitable parameter values
(cf.\ Eqs.\ (15) and (16) in Ref.\ \cite{10}). At the resonance
the coherence length is usually increased.

\subsection{Linear density profile}

The linear profile is by far the most important example,
because many actual profiles can be locally approximated
by it.
Let us parameterize the density as 
\begin{equation}
   N_{e}(x)=\lambda(\kappa -x)\ ,
\end{equation} 
where $\lambda$ and $\kappa$ are parameters. 
The adiabaticity condition, Eq.\ (\ref{33}), leads to
\begin{equation}
   \frac{(\Delta_{0}\sin 2\theta)^{2}}{E_0^{2}G_{F}\lambda}\gg 1\ . 
\label{36} \end{equation}
We first calculate 
\begin{equation}
 \int_{0}^{L}dx\Delta_{m}(x)
=\frac{1}{4c}[I(\kappa)-I(\kappa -L)]\ ,
\label{37}   
\end{equation}
with (see e.g.\ \cite{32}) 
\begin{equation}
I(x)=(2cx+b)\sqrt{a+bx+cx^{2}}
+\frac{4ac-b^{2}}{2\sqrt{c}}\ln(2\sqrt{c(a+bx+cx^{2})}+2cx+b)\
,
\end{equation}
and $a=\Delta_{0}^{2},\ b=-4\sqrt{2}E_0G_{F}\lambda\Delta_{0}
\cos 2\theta,\ c=8(E_0G_{F}\lambda)^{2}$. 
A direct substitution to Eq.\ (\ref{31}) 
then yields the effective oscillation length, in principle. 
Using Eqs.\ (\ref{31}), (\ref{41}) and
(\ref{37}), the effective coherence length could also be calculated,
but here that tedious calculation is omitted.

Since Eq.\ (\ref{37}) is not very illustrative, we will next consider the low
density limit (i.e.\ small $\lambda$). One has
\begin{equation}
     \int_{0}^{L}dx\Delta_{m}(x)=L\Delta_{0}+\sqrt{2}E_0 
G_{F}\lambda\cos 2\theta(L^{2}-2\kappa L)+O(\lambda^{2})
\end{equation}
yielding
\begin{equation}
  L^{osc}\approx\frac{4\pi E_0}
{\Delta_{0}}\left[1-\frac{\sqrt{2}E_0 G_{F}\lambda\cos
2\theta}{\Delta_{0}}(L-2\kappa)\right]\ ,
\end{equation}
where the second term in the brackets can be regarded as the first order
correction due to matter.
 The coherence length is to this order (with Eqs.\ (\ref{31}) and
(\ref{41}))
\begin{equation}
  L^{coh}\approx\frac{4\sqrt{2}\sigma_{x}E_{0}^{2}}{|\Delta_{0}|}\ ,
\end{equation}
i.e.\ the same as in vacuum. 
The lowest order correction of the coherence
length is in fact always of the second order: the linear term 
$\sim E_0 N_{e}(x)$
(see Eq.\ (\ref{32})) is wiped out by the ``operator'' 
$-1+E_0\partial_{E_0}$ in Eq.\ (\ref{41}).

\subsection{Exponential density profile}

Let the density of electrons be given by
\begin{equation}
    N_{e}(x)=\lambda e^{-\kappa x}\ ,   \label{45}
\end{equation}
where $\lambda$ and $\kappa$ are parameters. Eq.\ (\ref{33}) yields in this case
\begin{equation}
   \frac{\Delta_{0}\sin^{2}2\theta}{E_0 \cos 2\theta\kappa}\gg 1\ .  
\label{46} \end{equation}
Now we can write \cite{32}
\begin{equation}
   \int_{0}^{L}dx\Delta_{m}(x)=\frac{1}{\kappa}[I(1)-I(e^{-\kappa L})]\ ,
\end{equation}
where
\begin{eqnarray}
I(x)&=&
\sqrt{a+bx+cx^{2}}-
\sqrt{a}\ln\left(\frac{2a+bx+2\sqrt{a(a+bx+cx^{2})}}{x}\right)
\nonumber \\ && \mbox{ }
+\frac{b}{2\sqrt{c}}\ln(2\sqrt{c(a+bx+cx^{2})}+2cx+b)\
,
\end{eqnarray}
and $a=\Delta_{0}^{2},\ b=-4\sqrt{2}E_0 G_{F}\lambda\Delta_{0}
\cos 2\theta,\
c=8(E_0 G_{F}\lambda)^{2}$. Once again, this result with Eqs.\ 
(\ref{31}) and
(\ref{41}) would allow us to obtain the effective oscillation and coherence
lengths
in principle.

In the low density limit we can expand
\begin{equation}
    \int_{0}^{L}dx\Delta_{m}(x) = 
L\Delta_{0}+\frac{2\sqrt{2}E_0 G_{F}
\lambda\cos 2\theta}{\kappa}(e^{-\kappa
L}-1)-\frac{2(E_0 G_{F}\lambda\sin 2\theta)^{2}}
{\Delta_{0}\kappa}(e^{-2\kappa L}-1)
+ O(\lambda^3)\ .
\end{equation}
Now we get the first order correction to the effective oscillation length
\begin{equation}
   L^{osc}\approx\frac{4\pi E_0}{\Delta_{0}}
\left[1-\frac{2\sqrt{2}E_{0}G_{F}\lambda\cos
2\theta}{L\Delta_{0}\kappa}(e^{-\kappa L}-1)\right]\ .   \label{51}
\end{equation}
Eqs. (\ref{31}) and (\ref{41}) yield the correction to the 
effective coherence length
 \begin{equation}
L^{coh}\approx\frac{4\sqrt{2}\sigma_{x}E_0^{2}}{|\Delta_{0}|}
\left[1-\frac{2(E_0 G_{F}\lambda\sin
2\theta)^{2}}{L\Delta_{0}^{2}\kappa}(e^{-2\kappa L}-1)\right]\ ,
\label{53}
\end{equation}
which is  of the second order, in accordance with the remark made 
in the previous example.

We conclude with a simple numerical application on solar neutrinos. The electron
number density in the Sun is approximately \cite{33}
\begin{equation}
   N_{e}(x)=245N_{A}\exp\left(-10.54\frac{x}{R_{\odot}}\right)\mbox{cm}^{-3}\ ,
 \label{54}
\end{equation}
where $N_{A}$ is Avogadro's number. 
Hence (Eq.\ (\ref{45})) $\lambda\sim 10^{12}\
\mbox{eV}^{3}$ and $\kappa L\sim 10^{3}$ 
if neutrinos are produced in the 
center
of the Sun and detected on Earth. We use the values $\Delta_{0}\sim 10^{-4}\
\mbox{eV}^{2}$ and $E_{0}\sim 1\ \mbox{MeV}$, 
which fulfill the adiabaticity
condition, Eq.\ (\ref{46}), even 
for rather small values of $\theta$. Eq.\
(\ref{51}) yields the correction 
of the effective oscillation length
\begin{equation}
     \frac{E_0 G_{F}\lambda\cos 2\theta}{L\Delta_{0}\kappa}\sim 
10^{-4}\cos 2\theta\ ,
\end{equation}
where $G_{F}\sim 10^{-23}\ \mbox{eV}^{-2}$. 
The correction of the effective coherence length
is similarly
\begin{equation}
    \frac{(E_0 G_{F}\lambda\sin 2\theta)^{2}}
      {L\Delta_{0}^{2}\kappa}
\sim
10^{-5}\sin^{2}2\theta\ ,
\end{equation}
where Eq.\ (\ref{53}) was used. It is 
thus seen that the matter effect modifies
the oscillation and coherence lengths 
of solar neutrinos only slightly at least
within the framework of this example. 
This fact is, of course, due to the
insignificance of the density outside 
the Sun, i.e.\ the neutrinos propagate
mainly in vacuum between the Sun and Earth.
Similarly, it is to be believed that 
the matter effect is unimportant
for the oscillation of all extraterrestrial 
neutrinos (excluding, of course,
MSW effect and parametric resonance).

We finally call your attention to the fact 
that even though both the linear and
exponential density profiles violate the 
condition $|G_{F}N_{e}(x)|\ll E_{0}$ for
some
values of $x$, the sharpness of the wave 
packets forces $x$ to be situated in
harmless region (see Eqs.\ (\ref{10}) 
and (\ref{11})). It is due to this same
reason that the profile of Eq.\ (\ref{54}), 
written in spherical coordinates, is
directly applicable.
A more obvious reason is seen in the integrals in Eq.\
(\ref{31}).

\section{Nonadiabatic neutrino propagation in matter}

\subsection{Solution of the equation of motion for two neutrino flavors}

In the previous sections we have assumed that the neutrino propagation is
adiabatic. In most physical environments, however, the matter 
density may change so rapidly that the
nonadiabaticity must be taken into account.
When neutrinos propagate nonadiabatically the local matter ``eigenstates''
are not anymore eigenstates but they become mixed. This means that transitions
(i.e.\ level crossings) between the matter states may occur at certain 
probability. Mathematically this fact manifests
itself in the complete equation of motion (in the matter basis), which includes
also nondiagonal terms, whereas in the adiabatic limit the corresponding equation
is diagonal (cf.\ Eq.\ (\ref{3})).

In this section we look for a formal solution to the two flavor 
equation of motion
\cite{24,28,29,34,35}
\begin{equation}
   i\partial_{x}\left( \begin{array}{c} \psi_{1}(x,E) 
\\  \psi_{2}(x,E) 
\end{array}
\right)=\left( \begin{array}{cc}   -\frac{\Delta(x)}{4E} & -i\theta_{m}'(x) \\
                       i\theta_{m}'(x) & \frac{\Delta(x)}{4E} \end{array} \right)
\left( \begin{array}{c} \psi_{1}(x,E) \\  \psi_{2}(x,E) 
\end{array} \right)\ ,
\label{57}
\end{equation}
where $\Delta(x)\equiv\mu_{2}^{2}(x)-\mu_{1}^{2}(x)$ 
(this is not necessarily the same as in Eq.\ (\ref{32})
if the intermediate matter contains e.g.\ muons),
$\theta_{m}'(x)=\partial_{x}\theta_{m}(x)$ and $\theta_{m}(x)$ is the mixing angle
in matter. 
In the limit $4E|\theta_{m}'(x)|\ll\Delta(x)$, equivalent to the
previously considered adiabaticity condition, Eq.\ (\ref{33})
(for $\nu_e \leftrightarrow \nu_\mu$ oscillation), 
the nondiagonal terms can be neglected, and Eqs.\ (\ref{57}) and 
(\ref{3}) (with
$a=1,2$) correspond to each other perfectly (discarding terms
proportional to the identity matrix). 
Level crossings, on the other hand, are obviously caused by
the nondiagonal terms $\pm i\theta_{m}'(x)$. These 
issues, as well as exact calculations of the level crossing probabilities for
specific density profiles, have been extensively discussed by numerous authors;
see e.g.\ \cite{10},\cite{24}--\cite{29},\cite{34}--\cite{45}.

One should remember that Eq.\ (\ref{57}) describes the physical 
situation accurately enough 
if the following conditions hold (see e.g.\ Ref.\ \cite{22}):  
\begin{enumerate}
\item Neutrinos are relativistic.
\item The density is low enough, i.e. $|G_{F}N(x)|\ll E$, where $N(x)$ 
is the number
density of particle species relevant for a case under consideration 
 (mentioned already in the previous section for $N(x)=N_{e}(x)$).
\item The density must not change appreciably over a length scale equal to the
neutrino's de Broglie wavelength.
\end{enumerate}
The expression ``arbitrary density profile'', used in the following, is to be
understood in the context of the abovementioned limitations.

We now demonstrate how Eq.\ (\ref{57}) can be solved. Defining (we omit
here the energy dependence of the wave functions)
\begin{equation}
   \left( \begin{array}{c} \psi_{1}(x) \\  \psi_{2}(x) \end{array}
\right)=\frac{1}{\sqrt{2}}\left( \begin{array}{cc}  1  &  1  \\
         -1   & 1  \end{array} \right)  \left( \begin{array}{c} 
\phi_{1}(x) \\ \phi_{2}(x) \end{array} \right)   \label{58}
\end{equation}
one has
\begin{equation}
   i\partial_{x}\left( \begin{array}{c} \phi_{1}(x) \\  \phi_{2}(x) 
\end{array} \right)=\left( \begin{array}{cc}  0  & B(x)  \\
           B^{\ast}(x)  & 0  \end{array} \right)  \left( 
\begin{array}{c}
\phi_{1}(x) \\  \phi_{2}(x) \end{array} \right)\ , \label{59}
\end{equation}
where $B(x)\equiv -\frac{\Delta(x)}{4E}-i\theta_{m}'(x)$ (and $B^{\ast}(x)=
-\frac{\Delta(x)}{4E}+i\theta_{m}'(x)$). The solution of Eq. (\ref{59}) is
obtainable after some effort:
\begin{eqnarray} \label{60}
 \phi_1 (x) & = & C_1 \, \Xi_- (B,B^{\ast})
   + C_2 \, \Xi_+ (B,B^{\ast}) \ ,  \nonumber \\
 \phi_2 (x) & = & C_1 \, \Xi_- (B^{\ast},B)
   - C_2 \, \Xi_+ (B^{\ast},B) \ ,
\end{eqnarray}
where $C_1$ and $C_2$ are constants, and
\begin{equation} \label{neuf}
\Xi_{\pm} (Y,Z) \equiv 1 \pm i\int Y -\int Y\int Z \mp i\int Y\int
Z\int Y+\int Y\int Z\int Y\int Z \pm \cdots
\end{equation}
with e.g.\ $\int Y\int Z\int
Y\equiv\int_{x_{0}}^{x}dx_{1}Y(x_{1})\int_{x_{0}}^{x_{1}}
dx_{2}Z(x_{2})\int_{x_{0}}^{x_{2}}dx_{3}Y(x_{3})$.
Eq.\ (\ref{58}) yields finally 
\begin{eqnarray}   \label{61}
  \psi_{1}(x,E) & = & \frac{1}{2}\psi_{1}(x_{0},E) 
  (\Xi_- (B,B^{\ast})+\Xi_- (B^{\ast},B)) \nonumber \\
  &  &  -\frac{1}{2}\psi_{2}(x_{0},E) 
  (\Xi_+ (B,B^{\ast})-\Xi_+ (B^{\ast},B)) \ , \nonumber \\
 \psi_{2}(x,E) & = & \frac{1}{2}\psi_{1}(x_{0},E) 
   (-\Xi_- (B,B^{\ast})+\Xi_- (B^{\ast},B)) \nonumber \\
   &  &  +\frac{1}{2}\psi_{2}(x_{0},E) 
   (\Xi_+ (B,B^{\ast})+\Xi_+ (B^{\ast},B))\ .
\end{eqnarray}
A straightforward substitution shows that this is really the solution of Eq.\
(\ref{57}). As far as we know, no such complete solution has previously
been presented in the literature.
It is to be emphasized that Eq.\ (\ref{61}) applies to arbitrary
density profile, but on the other hand it has the undeniable deficiency of being
somewhat formal and perhaps not very useful for practical calculations. Further
discussion on the solution of Eq.\ (\ref{57}) and on the corresponding equation in
the flavor basis is found in Appendix B.

Let us show briefly that our solution gives a meaningful result in two 
specific cases. In
\emph{the adiabatic limit} $B(x)\approx -\frac{\Delta(x)}{4E}\approx B^{\ast}(x)$
and hence
\begin{equation}
   \psi_{1}(x,E)= \psi_{1}(x_{0},E) \, \Xi_- (B,B) =
\psi_{1}(x_{0},E)\exp\left(i\int_{x_{0}}^{x}dx'\frac{\Delta(x)}{4E}\right)\ ,
\end{equation}
and similarly
\begin{equation}
\psi_{2}(x,E)=
\psi_{2}(x_{0},E)\exp\left(-i\int_{x_{0}}^{x}dx'\frac{\Delta(x)}{4E}\right)\
, 
\end{equation}
i.e.\ the correct result is obtained. In \emph{the extremely nonadiabatic
limit}\footnote{Here
we consider the usual textbook example where the matter states
are related to electron and muon flavor states, and the matter 
does not contain muons. Remember that Eq.\ (\ref{57}) is not 
necessarily limited to this standard case.},
on the other hand, $B(x)\approx -i\theta_{m}'(x)\approx -B^{\ast}(x)$ (in the
resonance region), yielding
\begin{eqnarray}
   \psi_{1}(x,E) & = & \psi_{1}(x_{0},E)\left(1+\int B\int
B+\cdots\right)-\psi_{2}(x_{0},E)\left(i\int B+i\int B\int B\int B+\cdots\right)\
, \nonumber \\
   \psi_{2}(x,E) & = & \psi_{1}(x_{0},E)\left(i\int B+i\int B\int B\int
B+\cdots\right)+\psi_{2}(x_{0},E)\left(1+\int B\int B+\cdots\right)\ .
\end{eqnarray}
One may assume that $\theta_{m}(x)$ changes abruptly from $\pi/2$ to $\theta$ (the
vacuum mixing angle) in the resonance, i.e.
\begin{equation}
   \int_{x_{0}}^{x}B\approx -i\left(\theta-\frac{\pi}{2}\right)
\end{equation}
if $x_{0}(x)$ is before (after) the resonance. Hence
\begin{eqnarray}
  \psi_{1}(x,E) & = &
\psi_{1}(x_{0},E)\cos\left(\theta-\frac{\pi}{2}\right)
-\psi_{2}(x_{0},E)\sin\left(\theta-\frac{\pi}{2}\right)\
, \nonumber \\
  \psi_{2}(x,E) & = &
\psi_{1}(x_{0},E)\sin\left(\theta-\frac{\pi}{2}\right)
+\psi_{2}(x_{0},E)\cos\left(\theta-\frac{\pi}{2}\right)\ .
\end{eqnarray}
Putting e.g.\ $\psi_{1}(x_{0},E)=0,\ 
\psi_{2}(x_{0},E)=1$, we see that the level
crossing probability is
\begin{equation}
  |\psi_{1}(x,E)|^{2}=
\sin^{2}\left(\theta-\frac{\pi}{2}\right)=\cos^{2}\theta\ ,
\end{equation}
as it should in this limit \cite{29}.

\subsection{Level crossing probabilities and wave packets}

We now present how the existing results on the level crossing probabilities 
can be combined with the wave packet description in a
consistent manner. As in the previous section, we restrict to
two flavors which is sufficient for understanding the relevant
phenomena.

Consider a neutrino that propagates initially as $\psi_{2}$ and
has the usual Gaussian form (Eq.\ (\ref{6})), i.e.
\begin{equation}
    \psi_{1}(x_{0},E)=0\ \ \mbox{and}\ \
\psi_{2}(x_{0},E)=
N\exp\left[-\frac{(E-E_{2})^{2}}{4\sigma_{EP}^{2}}\right]\ ,
\end{equation}
where $N=(2\pi\sigma_{EP}^{2})^{-1/4}$. From Eq.\ (\ref{61}) it follows that 
\begin{eqnarray}
   \psi_{1}(x,E) & = &
-\frac{N}{2}\exp\left[-\frac{(E-E_{2})^{2}}{4\sigma_{EP}^{2}}\right]
f(x_{0},x,E)\ ,
\nonumber \\
   \psi_{2}(x,E) & = &
\frac{N}{2}\exp\left[-\frac{(E-E_{2})^{2}}{4\sigma_{EP}^{2}}\right]
h(x_{0},x,E)\ ,
\label{69}
\end{eqnarray}
where $f(x_{0},x,E)$ and $h(x_{0},x,E)$ 
represent the corresponding expansions in
Eq.\
(\ref{61}). Hence
\begin{equation}
   \psi_{1}(x,t)=-\frac{N}{2\sqrt{2\pi}}\int
dE\exp\left[-iEt-\frac{(E-E_{2})^{2}}{4\sigma_{EP}^{2}}\right]f(x_{0},x,E)\ ,
\end{equation}
where Eq. (\ref{2}) was used. Defining
\begin{equation}
\exp\left[-\frac{(E-E_{2})^{2}}{4\sigma_{EP}^{2}}\right]
f(x_{0},x,E) \equiv g(x_{0},x,E)
\end{equation}
one has
\begin{equation}
   \psi_{1}(x,t)=-\frac{N}{2}\hat{g}(x_{0},x,t)\ ,
\end{equation}
where the circumflex stands for the Fourier transform. The level crossing
probability is $|\psi_{1}(x,t)|^{2}$ but we integrate over time (cf.\ Sec.\
2) and have 
\begin{equation}
   |\psi_{1}(x)|^{2}=\int dt|\psi_{1}(x,t)|^{2}=\frac{N^{2}}{4}\int
dt|\hat{g}(x_{0},x,t)|^{2}=\frac{N^{2}}{4}\int dE|g(x_{0},x,E)|^{2}\ ,
\end{equation}
where Parseval's identity was used. 
We can express the level crossing probability for wave packets
using the respective probability for a plane wave, 
$P_{lc}(E)=\frac{1}{4}|f(x_{0},x,E)|^{2}$. Hence \begin{equation}
   P_{lc}(E_{2},\sigma_{EP})\equiv|\psi_{1}(x)|^{2}=N^{2}\int
dE\exp\left[-\frac{(E-E_{2})^{2}}{2\sigma_{EP}^{2}}\right]P_{lc}(E)\ ,  \label{74}
\end{equation}
where $P_{lc}(E_{2},\sigma_{EP})$ is the generalized level crossing
probability that takes into account the energy width of the wave packet.

The wave packet effects can be seen more clearly by expanding 
$P_{lc}(E)$ in series.
Assuming  $\sigma_{EP}$ to be small (Sec.\ 2), it is 
sufficient to take the lowest terms, and 
Eq.\ (\ref{74}) gives (with $E_{2}\rightarrow E$)
\begin{equation}
P_{lc}(E,\sigma_{EP})=P_{lc}(E)
+\frac{\sigma_{EP}^{2}}{2}\frac{\partial^{2}P_{lc}(E)}{\partial
E^{2}}+O\left(\sigma_{EP}^{4}\frac{\partial^{4}P_{lc}(E)}{\partial
E^{4}}\right)\ .   \label{76}
\end{equation}
This equation clearly indicates that the use of the wave packets modifies
the usual level crossing probabilities, $P_{lc}(E)$.
It also shows that the wave
packet correction is easily calculable if $P_{lc}(E)$ is known. 

Let us consider two  simple examples:
\begin{enumerate}
\item \emph{Linear density profile}. 
The well-known Landau-Zener probability is
\begin{equation}
    P_{LZ}(E)=\exp\left(-\frac{\pi}{4}Q\right)\ ,
\end{equation}
where $Q$, given in Eq.\ (\ref{33}),
should not be too small.
Using Eq.\ 
(\ref{76}) and
remembering that $Q\sim\frac{1}{E^{2}}$ (see Eq.\ (\ref{36})) one has
\begin{equation}
P_{LZ}(E,\sigma_{EP})=
\left[1+\left(\frac{\sigma_{EP}}{2E}\right)^{2}\left(\frac{\pi^{2}}{2}Q^{2}-3\pi
Q\right)\right]\exp\left(-\frac{\pi}{4}Q\right)\ .
\end{equation}
For the most relevant cases (i.e.\ $P_{LZ}(E)$ not too small) 
$(\pi Q)^{2}/2-3\pi Q$ is of the order of unity.

\item \emph{Exponential density profile} ($N_{e}(x)\sim e^{-\kappa x}$). Now
\cite{28,44}
\begin{equation}
   P_{lc}(E)=\frac{e^{-\pi\delta(1-\cos
2\theta)}-e^{-2\pi\delta}}{1-e^{-2\pi\delta}}=\frac{\sinh(B-A)}{\sinh B}e^{-A}\ ,
\end{equation}
where
\begin{equation}
  \delta=\frac{\Delta_{0}}{2E\kappa}\ ,\ \ \ A=\frac{\pi\delta}{2}(1-\cos
2\theta)\ ,\ \ \ B=\pi\delta\ .
\end{equation}
A tedious calculation gives
\begin{equation}
P_{lc}(E,\sigma_{EP})=
P_{lc}(E)\left[1+\left(\frac{\sigma_{EP}}{E}\right)^{2}\Gamma\right]\
,
\end{equation}
where
\begin{equation}
  \Gamma=A(1+\coth(B-A))(A-1+B(\coth B-1))+B(\coth(B-A)-\coth B)(1-B\coth B)\ .
\end{equation}
Again, for the interesting range of parameters $\Gamma \sim O(1)$ at most
when e.g.\ solar neutrinos are considered (with $E=1\ \mbox{MeV}$, 
$\Delta_{0}=10^{-4}\ \mbox{eV}^{2}$, and the density profile as given
in Eq.\ (\ref{54})).
\end{enumerate}
The wave packet
corrections turn out to be negligible
for small $\sigma_{EP}/E$. This happens
to be true in most physical environments,
e.g. for solar neutrinos
$\sigma_{EP}/E\sim
10^{-4}-10^{-5}$ \cite{12,24}.
If $\sigma_{EP}\sim E$, on the other hand,
the simple wave packet treatment is inaccurate \cite{24}. This 
fact is manifest also in our
calculation: if $\sigma_{EP}$ is not small enough, the integration in Eq.\
(\ref{74}) becomes problematic since $P_{lc}(E)$ is not necessarily
meaningfully defined for negative $E$ values.
Anyway, there might be at least in principle a situation where
the energy distribution of the neutrino wave function deviates
considerably from a plane wave. Our calculation suggests that
then the usual level crossing probabilities would not be 
totally reliable.

Finally, it is to be noted that the use of some specific $P_{lc}(E)$
restricts the location of $x_{0}$ and $x$ in Eqs. (\ref{61}) and (\ref{69});
$P_{LZ}(E)$, for example, is valid only if $x_{0}(x)$ is situated well before
(after) the resonance. If, on the contrary, $x_{0}$ and/or $x$ are/is in the
resonance region, $P_{LZ}(E)$ cannot be used.

\section{Summary}

We applied the wave packet formalism in order to study neutrino 
oscillations in matter
in the adiabatic limit, and found out that the effective 
oscillation and coherence lengths take into account the 
whole path the neutrino has traversed. Results for
the linear and exponential density profiles were briefly presented. 
The corrections for the predictions of observable 
fluxes of solar neutrinos seem to be quite small.
On the other hand, the matter may affect significantly the oscillation 
of the neutrinos e.g.\ in long baseline experiments for suitable
values of parameters.

We then considered the equation of motion for two neutrino flavors, and
managed to solve
it formally for arbitrary density profile. 
Our method clearly applies to any
differential equation of the same form. 

Finally, we showed that the level crossing
probabilities of the wave packets differ from those of plane waves. 
The difference is practically equivalent 
to a simple average over energy (cf.\ Eq.\ (\ref{74})),
not essentially related to the separation of wave packets.
The finite width of the wave packet does not result in
any observable effect for the physical situations we have considered.

We could have continued our work by combining the wave packet 
treatment of Sec.\ 2
with the complete solution of the two flavor equation in Sec.\ 4.1. 
In that case  the solution of Eq.\ (\ref{3}) (for $a=1,2$) should be 
replaced by Eq.\ (\ref{61}) in order to correctly take into account the
effects due to nonadiabaticity. 
However, the presented results suggest that this would not
reveal any new physics, so the calculation has been omitted.

In this work we used a model that
in principle is more physical and hence more accurate
than the models used normally. On the other hand,
since many of the calculations in this framework
are quite complicated, we considered the limits
of validity of the simpler plane wave approaches.
Our results show that the present neutrino observations
and the phenomena behind them can be described by a
plane wave model accurately enough, regarding the
current precision of the experiments.

\section*{Acknowledgements}

We are indebted to J. Maalampi for discussions and careful reading of the
manuscript. V.S. wishes to thank the Graduate School of Particle Physics
(Finland) for financial support.

\renewcommand{\theequation}{A\arabic{equation}}
\setcounter{equation}{0}

\appendix

\section*{Appendix A: Integration over $x$ in Eq.\ (\ref{12})}
We can write Eq.\ (\ref{12}) as
\begin{equation}
A_{\alpha \beta}(L,T) \sim  C(L,T) \int dx e^{F(x)}\ ,
\end{equation}
where
\begin{equation}
F(x)\equiv-\frac{v_{a}^{2}(0)}{4\sigma_{xP}^{2}}
\left(\int_{0}^{x}\frac{dx'}{v_{a}(x')}
-T\right)^{2}
-\frac{v_{a}^{2}(L)}{4\sigma_{xD}^{2}}
\left(\int_{L}^{x}\frac{dx'}{v_{a}(x')}\right)^{2}\ .
\end{equation}
The definition of the saddle point is in turn $F'(x_{0})=0\
(F'(x)=\frac{d}{dx}F(x))$, leading to
\begin{equation}
\sigma_{xD}^{2}\left(\int_{0}^{x_{0}}
\frac{dx'}{v_{a}(x')}-T\right)v_{a}^{2}(0)+\sigma_{xP}^{2}\int_{L}^{x_{0}}
\frac{dx'}{v_{a}(x')}v_{a}^{2}(L)=0\ .
\end{equation} From this one can solve that
\begin{equation}
\int_{0}^{x_{0}}\frac{dx'}{v_{a}(x')}=
\frac{\sigma_{xD}^{2}v_{a}^{2}(0)T+\sigma_{xP}^{2}v_{a}^{2}(L)\int_{0}^{L}
\frac{dx'}{v_{a}(x')}}{\sigma_{xD}^{2}v_{a}^{2}(0)+\sigma_{xP}^{2}v_{a}^{2}(L)}
\end{equation}
and  
\begin{equation}
\int_{L}^{x_0} \frac{dx'}{v_a (x')}=
\frac{ \sigma_{xD}^{2}v_{a}^{2}(0)
        \left( T-\int_0^L \frac{dx'}{v_a(x')} \right) }
{\sigma_{xD}^2 v_{a}^{2}(0)+\sigma_{xP}^2 v_{a}^{2}(L)}
\ ,
\end{equation}
and finally after some algebra
\begin{equation}
F(x_{0})=
-\frac{1}{4}\frac{\left(T-\int_{0}^{L}\frac{dx}{v_{a}(x)}\right)^{2}}
{\left(\frac{\sigma_{xP}}{v_{a}(0)}\right)^2+
\left(\frac{\sigma_{xD}}{v_{a}(L)}\right)^2}
\ .           \label{17}
\end{equation}

Let us now examine the higher derivatives of $F(x)$. 
The second derivative is
\begin{eqnarray}
    F''(x) & = &
-\frac{1}{2\sigma_{xP}^{2}}
\left(\frac{v_{a}(0)}{v_{a}(x)}\right)^{2}+
\frac{1}{2\sigma_{xP}^{2}}
\left(\int_{0}^{x}\frac{dx'}{v_{a}(x')}-T\right)
\left(\frac{v_{a}(0)}{v_{a}(x)}\right)^{2}v_{a}'(x)
\nonumber  \\
  &  &
-\frac{1}{2\sigma_{xD}^{2}}\left(\frac{v_{a}(L)}{v_{a}(x)}\right)^{2}
+\frac{1}{2\sigma_{xD}^{2}}\left(\int_{L}^{x}\frac{dx'}{v_{a}(x')}\right)
\left(\frac{v_{a}(L)}{v_{a}(x)}\right)^{2}v_{a}'(x)\
.  \label{18}
\end{eqnarray}
In the relativistic limit $v_{a}(x)= 1+O(\mu^{2}_{a}/E^{2})$, and 
Eq.\
(\ref{18}) reduces to\footnote{The adiabaticity 
further reinforces the smallness
of $v_{a}'(x)$.} 
\begin{equation}
F''(x_{0}) = -\frac{1}{2\sigma_{xP}^{2}}
-\frac{1}{2\sigma_{xD}^{2}}+O\left(\frac{L-T}{\sigma_{xP,D}^{2}}
\frac{d}{dx_{0}}\frac{\mu_{a}^{2}(x_{0})}{E^{2}}\right)\ ,
\end{equation}
where effectively $\sigma_{xP,D}^{2}=\max \{\sigma_{xP}^{2},
\sigma_{xD}^{2}\}$. From Eq.\ (\ref{18}) 
one sees also that the third derivative
of $F(x)$ includes terms proportional 
to $v_{a}'(x),\ (v_{a}'(x))^{2}$ and
$v_{a}''(x)$. It is thus obvious that 
$F'''(x)$ and all the higher derivatives
are negligible in the relativistic limit, 
and we can approximate (with
$F'(x_{0})=0$)
\begin{equation}
     F(x)\approx F(x_{0})+\frac{1}{2}F''(x_{0})(x-x_{0})^{2}\ .
\end{equation}
The relevant part of the integral of Eq.\ (\ref{12}) becomes
\begin{equation}
     \int dx e^{F(x)}
\approx
e^{F(x_{0})} \int
dx\exp\left[-\frac{1}{4}\left(\frac{1}{\sigma_{xP}^{2}}+
\frac{1}{\sigma_{xD}^{2}}\right)(x-x_{0})^{2}\right]\ ,  
\label{uusi}
\end{equation}
where the final Gaussian integral gives only 
an unimportant numerical factor.

\renewcommand{\theequation}{B\arabic{equation}}
\setcounter{equation}{0}
\section*{Appendix B: Additional calculations in connection with the two 
flavor equation of motion}

We present another way of solving Eq.\ (\ref{57}). Instead of Eq.\ (\ref{58}), we
now define (compare to Ref.\ \cite{43})
\begin{equation}
   \left( \begin{array}{c} \psi_{1}(x) \\  \psi_{2}(x) \end{array} \right)=\left(
\begin{array}{cc}   \exp\left(i\int_{x_{0}}^{x}dx'\frac{\Delta(x')}{4E}\right) & 0
\\
   0   &  \exp\left(-i\int_{x_{0}}^{x}dx'\frac{\Delta(x')}{4E}\right) \end{array}
\right)
 \left( \begin{array}{c} \phi_{1}(x) \\  \phi_{2}(x) \end{array} \right)\ , 
\end{equation}
leading to
\begin{equation}
   i\partial_{x}\left( \begin{array}{c} \phi_{1}(x) \\  \phi_{2}(x) \end{array}
\right)=\left( \begin{array}{cc}  0  & -iD(x)  \\
               iD^{\ast}(x)  & 0  \end{array} \right)  \left( \begin{array}{c}
\phi_{1}(x) \\  \phi_{2}(x) \end{array} \right)\ , \label{a2}
\end{equation}
where
\begin{equation}
   D(x)\equiv\theta_{m}'(x)
\exp\left(-i\int_{x_{0}}^{x}dx'\frac{\Delta(x')}{2E}\right)\ .
\end{equation}
Eq.\ (\ref{a2}) yields easily
\begin{eqnarray}   
 \phi_{1}(x) & = & C_{1}\left(1-\int D-\int D\int D^{\ast}+\int D\int D^{\ast}\int
D+\int D\int D^{\ast}\int D\int D^{\ast}+\cdots\right) \nonumber \\
  &  &  +C_{2}\left(1+\int D-\int D\int D^{\ast}-\int D\int D^{\ast}\int D+\int
D\int D^{\ast}\int D\int D^{\ast}+\cdots\right)\ ,  \nonumber \\
  \phi_{2}(x) & = & C_{1}\left(1+\int D^{\ast}-\int D^{\ast}\int D-\int
D^{\ast}\int D\int D^{\ast}+\int D^{\ast}\int D\int D^{\ast}\int D+\cdots\right)
      \\
  &  &  -C_{2}\left(1-\int D^{\ast}-\int D^{\ast}\int D+\int D^{\ast}\int D\int
D^{\ast}+\int D^{\ast}\int D\int D^{\ast}\int D+\cdots\right)\ , \nonumber 
\end{eqnarray}
where the notation is as before (see Eq.\ (\ref{neuf})). Finally one has
\begin{eqnarray}
  \psi_{1}(x) & = & \exp\left(i\int_{x_{0}}^{x}dx'\frac{\Delta(x')}{4E}\right)
                    \nonumber  \\
  &  & \times\left[\psi_{1}(x_{0})\left(1-\int D\int D^{\ast}+\int D\int
D^{\ast}\int D\int D^{\ast}+\cdots\right)\right. \nonumber  \\
  &  & \left.\ \ \ +\psi_{2}(x_{0})\left(-\int D+\int D\int D^{\ast}\int
D+\cdots\right)\right]\ ,  \nonumber  \\
   \psi_{2}(x) & = & \exp\left(-i\int_{x_{0}}^{x}dx'\frac{\Delta(x')}{4E}\right)
                    \nonumber  \\
  &  & \times\left[\psi_{1}(x_{0})\left(\int D^{\ast}-\int D^{\ast}\int D\int
D^{\ast}+\cdots\right)\right. \nonumber  \\
  &  & \left.\ \ \ +\psi_{2}(x_{0})\left(1-\int D^{\ast}\int D+\int D^{\ast}\int
D\int D^{\ast}\int D+\cdots\right)\right]\ ,
\end{eqnarray}
where the energy dependence is not written down explicitly. We thus see that Eq.\
(\ref{61}) is not the only form in which the solution of Eq.\ (\ref{57}) can be
expressed.              

The equation of motion in the flavor basis is \cite{24,28} (and many others)
\begin{equation}
   i\partial_{x}\left( \begin{array}{c} \psi_{e}(x) \\  \psi_{\mu}(x) \end{array}
\right)=\frac{1}{4E}\left( \begin{array}{cc}  -\Delta_{0}\cos 2\theta +A_{c}(x)  &
\Delta_{0}\sin 2\theta  \\
     \Delta_{0}\sin 2\theta  &  \Delta_{0}\cos 2\theta -A_{c}(x)  \end{array}
\right)  \left( \begin{array}{c} \psi_{e}(x) \\  \psi_{\mu}(x) \end{array}
\right)\ , 
\end{equation}
where $A_{c}(x)=2\sqrt{2}EG_{F}N_{e}(x)$ and other notations are obvious. Since
the calculation proceeds as above, we omit the details and content ourselves with
giving the final answer:
\begin{eqnarray}
  \psi_{e}(x) & = & \exp\left[\frac{i}{4E}\int_{x_{0}}^{x}dx'(\Delta_{0}\cos
2\theta -A_{c}(x'))\right]  \nonumber  \\
  &  & \times\left[\psi_{e}(x_{0})\left(1-\int F\int F^{\ast}+\int F\int
F^{\ast}\int F\int F^{\ast}+\cdots\right)\right. \nonumber  \\
  &  & \left.\ \ \ +\psi_{\mu}(x_{0})\left(-i\int F+i\int F\int F^{\ast}\int
F+\cdots\right)\right]\ ,  \nonumber  \\
   \psi_{\mu}(x) & = & \exp\left[\frac{-i}{4E}\int_{x_{0}}^{x}dx'(\Delta_{0}\cos
2\theta -A_{c}(x'))\right]  \nonumber  \\
  &  & \times\left[\psi_{e}(x_{0})\left(-i\int F^{\ast}+i\int F^{\ast}\int F\int
F^{\ast}+\cdots\right)\right. \nonumber  \\
  &  & \left.\ \ \ +\psi_{\mu}(x_{0})\left(1-\int F^{\ast}\int F+\int F^{\ast}\int
F\int F^{\ast}\int F+\cdots\right)\right]\ ,
\end{eqnarray}
where
\begin{equation}
  F(x)\equiv\frac{\Delta_{0}\sin
2\theta}{4E}\exp\left[\frac{-i}{2E}\int_{x_{0}}^{x}dx'(\Delta_{0}\cos 2\theta
-A_{c}(x'))\right]\ .
\end{equation}


\newpage
\begin{thebibliography}{123}
\bibitem{5}Y. Fukuda et al., Phys.\ Rev.\ Lett.\ \textbf{81}, 1562 (1998).
\bibitem{1}B. Pontecorvo, Zh.\ Eksp.\ Theor.\ Fiz.\ \textbf{33}, 549 (1957) [JETP
\textbf{6}, 429 (1958)].
\bibitem{2}Z. Maki, M. Nakagawa and S. Sakata, Prog.\ Theor.\ Phys.\ \textbf{28},
870 (1962).
\bibitem{3}B. Pontecorvo, Zh.\ Eksp.\ Theor.\ Fiz.\ \textbf{53}, 1717 (1969) [JETP
\textbf{26}, 984 (1968)].
\bibitem{4}S. M. Bilenky and B. Pontecorvo, Phys.\ Rep.\ \textbf{41}, 225 (1978).

\bibitem{6}C. Giunti, C. W. Kim and U. W. Lee, Phys.\ Rev.\ D \textbf{44}, 3635
(1991).
\bibitem{7}J. Rich, Phys.\ Rev.\ D \textbf{48}, 4318 (1993).
\bibitem{8}S. M. Bilenky and S. T. Petcov, Rev.\ Mod.\ Phys.\ \textbf{59}, 671
(1987).
\bibitem{9}B. Kayser, Phys.\ Rev.\ D \textbf{24}, 110 (1981).

\bibitem{10}C. Giunti, C. W. Kim and U. W. Lee, Phys.\ Lett.\ B \textbf{274}, 87
(1992).
\bibitem{11}C. Giunti, C. W. Kim, J. A. Lee and U. W. Lee, Phys.\ Rev.\ D
\textbf{48}, 4310 (1993).
\bibitem{12}K. Kiers, S. Nussinov and N. Weiss, Phys.\ Rev.\ D \textbf{53}, 537
(1996).
\bibitem{13}C. Giunti, C. W. Kim and U. W. Lee, Phys.\ Lett.\ B \textbf{421}, 237
(1998).
\bibitem{14}C. Giunti and C. W. Kim, Phys.\ Rev.\ D \textbf{58}, 017301 (1998). 
\bibitem{15}M. Nauenberg, Phys.\ Lett.\ B \textbf{447}, 23 (1999).
\bibitem{17}W. Grimus and P. Stockinger, Phys.\ Rev.\ D \textbf{54}, 3414 (1996).
\bibitem{19}W. Grimus, P. Stockinger and S. Mohanty, Phys.\ Rev.\ D \textbf{59},
013011 (1998).
\bibitem{20}L. Stodolsky, Phys.\ Rev.\ D \textbf{58}, 036006 (1998).
\bibitem{16}A. Widom and Y. N. Srivastava, hep-ph/9608476.
\bibitem{18}Y. V. Shtanov, Phys.\ Rev.\ D \textbf{57}, 4418 (1998).
\bibitem{21}L. Wolfenstein, Phys.\ Rev.\ D \textbf{17}, 2369 (1978).
\bibitem{22}A. Halprin, Phys.\ Rev.\ D \textbf{34}, 3462 (1986).
\bibitem{23}P. D. Mannheim, Phys.\ Rev.\ D \textbf{37}, 1935 (1988).
\bibitem{24}C. W. Kim and A. Pevsner, \emph{Neutrinos in Physics and
 Astrophysics} (Harwood Academic, Chur, 1993).
\bibitem{25}D. N\"{o}tzold, Phys.\ Rev.\ D \textbf{36}, 1625 (1987).
\bibitem{26}S. Toshev, Phys.\ Lett.\ B \textbf{196}, 170 (1987).
\bibitem{27}S. T. Petcov, Phys.\ Lett.\ B \textbf{200}, 373 (1988).
\bibitem{28}A. B. Balantekin, Phys.\ Rev.\ D \textbf{58}, 013001 (1998).
\bibitem{29}C. W. Kim, S. Nussinov and W. K. Sze, Phys.\ Lett.\ B \textbf{184},
403 (1987).
\bibitem{30}C. Giunti, C. W. Kim and U. W. Lee, Phys.\ Rev.\ D \textbf{45}, 2414
(1992).
\bibitem{31}S. P. Mikheyev and A.Y. Smirnov, Sov.\ J.\ Nucl.\ Phys.\ \textbf{42},
913 (1985); JETP \textbf{64}, 4 (1986).
\bibitem{32}I. S. Gradshteyn and I. M. Ryzhik, \emph{Tables of Integrals, 
Series and Products} (Academic Press, Orlando, 1980).
\bibitem{33}J. N. Bahcall and R. K. Ulrich, Rev.\ Mod.\ Phys.\ \textbf{60}, 297
(1988).
\bibitem{34}T. K. Kuo and J. Pantaleone, Phys.\ Rev.\ D \textbf{39}, 1930 (1989).
\bibitem{35}A. B. Balantekin and J. F. Beacom, Phys.\ Rev.\ D \textbf{54}, 6323
(1996).
\bibitem{36}L. Landau, Phys.\ Z.\ Sovjetunion \textbf{2}, 46 (1932).
\bibitem{37}C. Zener, Proc.\ R.\ Soc.\ London, Ser.\ A \textbf{137}, 696 (1932).
\bibitem{38}S. P. Rosen and J. M. Gelb, Phys.\ Rev.\ D \textbf{34}, 969 (1986).
\bibitem{39}V. Barger, R. J. N. Phillips and K. Whisnant, Phys.\ Rev.\ D
\textbf{34}, 980 (1986).
\bibitem{40}S. J. Parke, Phys.\ Rev.\ Lett.\ \textbf{57}, 1275 (1986).

\bibitem{41}P. Pizzochero, Phys.\ Rev.\ D \textbf{36}, 2293 (1987).
\bibitem{42}A. B. Balantekin, S. H. Fricke and P. J. Hatchell, Phys.\ Rev.\ D
\textbf{38}, 935 (1988).
\bibitem{43}J. C. D'Olivo, Phys.\ Rev.\ D \textbf{45}, 924 (1992).
\bibitem{44}M. Bruggen, W. C. Haxton and Y. Z. Qian, Phys.\ Rev.\ D \textbf{51},
4028 (1995).
\bibitem{45}A. B. Balantekin, J. F. Beacom and J. M. Fetter, Phys.\ Lett.\ B
\textbf{427}, 317 (1998).

\end{thebibliography}
\end{document}